# Preconceptual Modeling in Software Engineering: Metaphysics of Diagrammatic Representations

Sabah Al-Fedaghi[*]

*Computer Engineering Department*
*Kuwait University*
*Kuwait*
salfedaghi@yahoo.com, sabah.alfedaghi@ku.edu.kw

*Abstract* – Conceptual modeling of a portion of the world is a necessary prerequisite to set the stage and define software system boundaries. In this context, one of the challenges is to provide a unified framework to create a comprehensive representation of the targeted domain. According to many researchers, conceptual model (CM) development is a hard task, and system requirements are difficult to collect, causing many miscommunication problems. Accordingly, CMs require more than modeling ability alone: they first require an understanding of the targeted domain that the model attempts to represent. Accordingly, a preconceptual modeling (pre-CM) stage is intended to address ontological issues before typical CM development is initiated. It involves defining a portion of reality when entities and processes are differentiated and integrated as unified wholes. This pre-CM phase forms the focus of research in this paper. The purpose is not show *how to model*; rather, it is to demonstrate *how to establish a metaphysical basis of the involved portion of reality*. To demonstrate such a venture, we employed the so-called thinging machine (TM) modeling that has been proposed as a high-level CM. A TM model integrates staticity and dynamism grounded in a fundamental construct called a *thimac* (things/machine). It involves two modes of reality, *existence* (events) and *subsistence* (regions: roughly, specifications of things and processes). Currently, the dominant approach in CM has evolved to limit its scope of application to develop ontological categorization (types of things). In contrast, advocates of TM modeling have pursued a broader metaphysical study of the nature of the domain's things and processes beyond categorization. In the TM approach, pre-CM metaphysics is viewed as a part and parcel of CM itself. The general research problem is how to map TM constructs to *what is out there in the targeted domain*. Discussions involve the nature of thimacs (things and processes) and subsistence and existence as they are *superimposed* over each other in reality. Specifically, we make two claims, (i) the perceptibility of regions as a phenomenon and (ii) the distinctiveness of existence as a construct for events. The results contribute to further the understanding of TM modeling in addition to introducing some metaphysical insights.

## I. INTRODUCTION

We propose in this paper to view software engineering as a spectrum that extends across preconceptual modeling (pre-CM), conceptual modeling (CM), systems design, and implementation stages (Fig. 1).

----------------------

[*] Retired June 2021, seconded fall semester 2021/2022

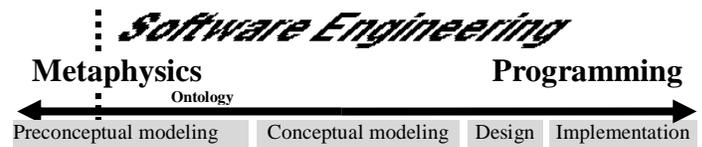

Fig. 1 Stages of development in software engineering.

The pre-CM stage is intended to address metaphysical issues before typical CM starts. It involves defining a portion of reality when entities and processes are differentiated from their surroundings and integrated as unified wholes to define the system boundaries.

In software engineering, CMs are widely used to build a high-level representation of some portion of reality in the early phases of systems development when requirements are being specified. The pre-CM involves defining a portion of reality elements such as *entity*, *class*, *object*, *relationship*, *time* and *dynamism*.

### A. Problem and Aim

According to many researchers, CM development is a hard task, and system requirements are difficult to collect, causing many miscommunication problems. One of the methodological challenges is to provide a unified framework to create a comprehensive yet insightful and understandable representation. For example, according to [1], modeling a portion of reality as a necessary prerequisite to define the system boundaries has become problematic. CMs are hard to understand for somebody not involved in the model definition [2]. Requirement elicitation needs stakeholders' viewpoints, which are challenging to collect, causing additional miscommunication issues [3]. This phase, which involves identifying systems' requirements, has become the hardest stage to learn because of its theoretical nature and the diversity of the knowledge required [3]. In abstracting from real-world situations, various stakeholders would interpret the observed situations differently; therefore, dissimilar perceptions and opinions on reality might lead to misunderstandings and distorted review results.

Accordingly, CMs require more than modeling ability alone: they first require the developer to understand the target domain that the model attempts to represent. The pre-CM phase also requires identifying objects and processes and their natural boundaries through a representational structure. This makes the phase of special importance, targeting not the *modeling itself* but rather *forming a metaphysical basis of the CM*.

## B. Proposed Solution

To demonstrate how to achieve such a goal, we utilized the so-called thinging machine (TM), which has been proposed as a high-level CM model. Currently, the dominant approach in CM, the Unified Modeling Language (UML), has evolved to pursue metaphysics as being "outside of its boundaries," limiting its scope of application to develop ontological categorization (types of things) in CM (e.g., Bunge-Wand-Weber ontology) [4]. Such an approach uses ontology in CM while denying its proximity to metaphysics. This seems to be a severe restriction, since the modeler in CM has a direct connection to (natural) reality, in contrast to physics-like sciences, which typically prioritize *what is objective* over the human understanding. Recently, there have been proposals to develop a more general approach (e.g., what is called a universal conceptual modeling language) [5].

Instead, the TM modeling in this paper is used to pursue a broader metaphysical exploration of the nature of modeled things in contrast to limiting the goal to categorization. The general research problem is how to map TM constructs to *what is out there in the targeted domain*. The purpose is not show *how to model*; rather, it is to demonstrate *how to form a metaphysical basis of the model* to provide a unified framework for conceptualization. This task is difficult because the process involves ineliminable and imprecise terms with various interpretations, perceptions and opinions on reality that might lead to distorted conceptualizations. The conceptualization process parallels the established models of scientific understanding as well as various descriptions of fundamental entities and phenomena, models of which have since risen and fallen.

Such research may also benefit metaphysics by introducing a new tool (diagrammatic language) to analyze metaphysical notions. Reference [6] proposed that metaphysics consist of "the generalisation and unification of concepts, principles and even whole theories of single disciplines to abstract and overarching concepts, principles and theories – asks also for the application of aprioristic methods such as conceptual analysis." The TM modeling language can serve as an apparatus for progress in this direction.

## C. Preconceptual Modeling and Preconceptualization

The notion of preconceptualization is said to have been coined by Heidegger [3], referring to a previous knowledge about a concept. Piaget, in his Stage Theory, distinguishes a preconceptual stage with regard to the understanding of class membership and internal representations. In software development, analysts investigate to find the ideas behind the discourse and depict such studies as *preconcepts* [3].

Additionally, preconceptualization is used in software engineering in the so-called *preconceptual schema* as an intermediate stage between natural language and UML conceptual Schemas. Preconceptual schemas have been used to represent concepts' relationships belonging to a certain domain and to search for the automated generation of UML conceptual schemas. Reference [7] discussed preconceptual schemas that are used for generating *conceptual diagrams*. Lately, [8] proposed an approach to enabling computers to interpret basic human-made preconceptual schemas.

Knowledge representation has been applied in software engineering to tasks such as requirements elicitation, and formal specification. [3]. Several paradigms have been used in this context, such as semantic networks, frames, production rules and predicate logic to specify preconceptual schemas and transform them to UML diagrams.

Unlike these preconceptualization ideas, pre-CM is mainly a premodeling conceptualization process. That is, it is conceptualization that targets modeling. Preconceptualization used in preconceptual schemas research is a type of CM itself, with purposes such as generating diagrams and source code of a given domain as well as employing linguistic definitions to get stakeholder understanding and validation by using a controlled language, which is closer to the stakeholder discourse [7].

## D. Content of Paper

The next section includes an enhanced description of the TM model as a world of *thimacs*. Subsections are devoted to discussing what a TM thing and a TM machine are. TM modeling has been applied in many research areas (e.g., [9-11]); hence, the examples in Section 3 apply it in new areas. Accordingly, Section 3 presents thee new examples:

- The first example involves hierarchical representation to organize concepts in terms of such relations as *is_a* and *has_a* (taken from [12]).
- The second example models a graphical information system (GIS) that involves representing locations and defining operations related to spatial objects (taken from [13]).
- The third example involves Zeno's paradox of motion of continuity and change, which is given by observing an arrow that has been shot from the bow. This example introduces TM modeling of the notion of movement.

Sections 4, 5 and 6 include the subject of TM metaphysics. Section 4 involves discussions about the nature of *thingness* and *thimacness* and explains how subsistence and existence are *superimposed* over each other in reality. Sections 5 and 6 center on the subsistence of *regions* and the existence of *events*, respectively. Specifically, in these sections, we make two claims:

(i) The perceptibility of regions as a phenomenon and
(ii) The distinctiveness of pure existence (called exicons) from ordinary existence (events).

## II.    THE TM MODEL

The TM model is a one-category CM based on the notion of TM, which integrates **staticity** and **dynamism** grounded on a fundamental construct called *thimac* (things/machines). Bold terms in this text will be defined later, but for now, intuitive common understanding is sufficient. The thimac has a dual mode of **reality**: the machine side and the thing (illustrated in Fig. 2). Accordingly, a thimac may be referred to as a *thing* or as a *machine*. As shown in Fig. 2, a thimac is defined in terms of five generic actions: *create*, *process*, *release*, *transfer* and *receive*. A thimac is a *machine* when it acts on other thimacs, and it is a *thing* when it is the object of actions by other thimacs.

A thimac may be constructed from a subset of these five generic actions (also called generic **regions**). A *create* action is mandatory for the existence of any thimac. A thing is what can be created, processed, released, transferred and/or received. A machine is defined in terms of the generic action according to the structure shown in Fig. 3. Generic actions appear only within thimacs, according to the TM structure in Fig. 3. For simplicity's sake, we assume that things are always *accepted* at their destination (see Fig. 3); therefore, we focus on the indicated actions above.

### A.    Thimacs

A thimac is a unit (indivisible) that cannot be disintegrated into its generic action (create, process, etc.), but it may have parts designated as subthimacs. Thimacs form a network of things that articulate the underlying structure of the targeted domain. They are the wholes (entities, e.g., a stone; and processes, e.g., traffic) of world view composition that can be built from subthimacs, subsubthimacs and so on. Each thimac is woven from and in other thimacs, forming an organized whole. According to such a view, the whole holds together as one thimac and is more than a mere entanglement of interconnections of similar thimacs. The thimac as a machine is the basic unit of the whole constructed as a repeated crystal lattice structure.

Every thimac is distinct from every other thimac according to its superthimacs (analogous to uniqueness in classical space) or subthimacs (analogous to *classical definitions*). In a classical view, a definition provides characteristics that are necessary and jointly sufficient for membership in the category [14].

Some subthimacs may be actualized without the whole thimac being actualized. On the other hand, a whole thimac may not be actualized without the **actualization** of some of its subthimacs (e.g., a person cannot exist without a [some type of] heart). Thimacs may attach themselves to other thimacs and form new thimacs.

The world is divisible in a complexity of overlapping thimacs, but the thimacs themselves are formed (not divisible) in terms of (not necessarily all) generic actions of the TM. To clarify this notion of the relationship between thimacs and action, we can use the set theory version of integers, where sets include {}, {1}, … but 1 by itself is not a set. Similarly, in a TM, a **named** *create* is a thimac, but *create* by itself is not a thimac.

The thimac as a machine is irreducible and holds its actions together such that none can separate from the machine. Ontologically, machine covers the **thingness**, and as will be seen later, **nonmachine-ness** is **nothingness**. Accordingly, in a TM, *reality* in itself is every*thing* (thimacs), but there is *outside* that is not in reality. The outside has no create, process, release, transfer or receive actions.

Additionally, the TM model includes *storage* and *triggering* (denoted by a dashed arrow in this paper's figures), which initiate a flow from one machine to another. Triggering is a transformation from one series of movements to another (e.g., electricity triggers the generation of heat).

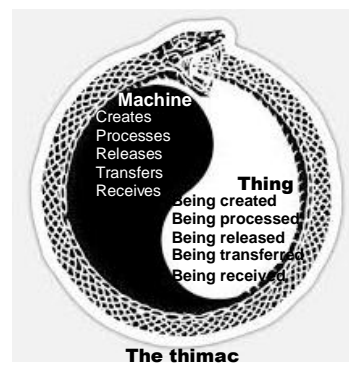

**Fig. 2 The dual nature of the thimac – Original image from Spreadshirt.**
https://www.spreadshirt.com/shop/design/ouroboros+yin+yang+spiritual+meditation+gift+sticker-D618s50f17b4bb016ae63572ff?sellable=jwpXOywmr5Ud7z7gGg9M-1459-215

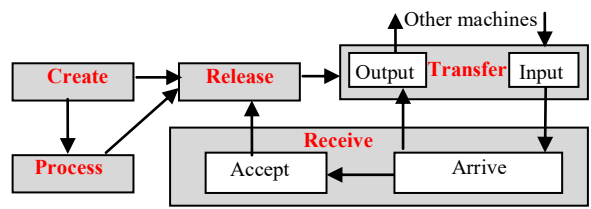

**Fig. 3 Thinging machine.**

### B.    The Thing

Things are known by their names, and further naming (higher level of naming) is related to these named things [15]. As we will discuss later, these names have meanings specified by diagrammatic TM regions that describe things and processes of various scales and complexity.

All things can flow (i.e., things that flow are things that can be created, released, transferred, received and processed). A TM model is a conceptual map that shows the propagation and progression of things. A thing flows from a machine through its *release* and *transfer* (output) to another machine, reaching this second machine's transfer (input) to settle in its receive. Flow refers to transformation in the situations of a thing among creation, procession, releasing, transferring and receiving.

## C. The Machine

Deleuze stated that "everything is a machine," and hence, "from this point of view natural substances and artificial constructions, candelabras and trees, turbine and sun are no longer any different" [16]. In a TM, instead of everything being a machine, everything has a dual model of reality – the thing mode and the machine mode. To define things as machines is to *create* the Fig. 3 blueprint with a subset of {process, release, transfer, receive} actions. The TM machine is the intricate ontological structure of thimacs.

In TM modeling, the thimac *machine* executes five actions: *create*, *process*, *release*, *transfer* and *receive*. Each of these static (outside time) action is a capacity to act and becomes a *generic event* when merged with time. Thimacs are realized by creating, processing, releasing, transferring and/or receiving thimacs. Each thimac is affected by/affects the thimacs in contact with it through releasing, transferring and receiving. In this machine world, there is no meaningful distinction between subject and object.

A thimac's actions, as shown in Fig. 3, are described as follows.

1) *Arrive:* A thing arrives to a machine.
2) *Accept:* A thing enters the machine. For simplification, the arriving things are assumed to be *accepted* (see Fig. 3); therefore, *arrive* and *accept* combine actions into the *receive* action. Thus, the thing becomes inside the machine.
3) *Release:* A thing is ready for transfer outside the machine. It is dismissed from the interior of the machine, waiting to be shipped out.
4) *Process:* A thing is changed, handled and examined, but no new thing is generated.
5) *Transfer:* A thing crosses a boundary as input into or output from a machine.
6) *Create:* A new thing manifests in a machine. The creation action underlies all existing things in reality.

The main claim concerning the TM is that five-action thimacs are the fundamental constituents out of which everything in reality is composed, all the way down and all the way up. These five actions are called generic (independent) actions. Martin Heidegger [17] encourages further research on "generic processes" applied to a thing. According to [18], recent years have seen renewed interest in the semantics of generics and their various types of appeal.

As we will discuss later (Example 3, Section 3), a changing thing flowing inside each generic action takes a continuous form (does not have any gaps or holes) as a generic event. Therefore, change is simply applied to different generic actions by a thing at different times. The notion of continuity seems to presume that the continuous TM generic action forms a basis by which the change continues while the involved (flowing) thing remains the same (e.g., continuous *transfer* of that same thing to reach the end of the transferring action).

## III. TM MODELING EXAMPLES

### A. Example 1: Hierarchical Representation

Reference [12] describes a certain sentence (e.g., *A canary can fly*). People store only the generalization that birds can fly and infer that a canary can fly from the stored information that a canary is a bird and birds can fly. The organization is illustrated by a memory structure of a three-level hierarchy (Fig. 4). For [12], hierarchical model is perhaps the simplest and most natural method to organize concepts. A predefined *is_a* relation organizes natural objects in the form of a tree. Other relations (e.g., *can_a*, *has_a*) are useful for building additional associations between nodes. The *is_a* relation introduces inheritance with properties of the concepts from higher levels of the taxonomy [12]. Although hierarchical models have many applications, it is clear that the memory structures are not static. Some relations among concepts cannot be accommodated in the hierarchical model [12].

From the TM model point of view, this type of depiction lacks the representation of *behavior*. For our purpose, it is sufficient to use a partial view of this organization, as shown in Fig. 4. Fig. 5 shows the corresponding TM static model that preserves the hierarchy of information and enhances the behavioral representation. Staticity exemplifies the limited world of Fig. 5 as potentialities:

- *There are* canaries, ostriches and sharks (dotted ellipses 1, 2 and 3) as individuals and groups.
- Canaries fly in the sky (flow from what is assumed initially) to sky (4): actions *release*, *transfer* (out), *transfer* (in), *receive*.
- Sharks attack humans (5).
- Etc.

Note that, for simplicity's sake, creation is not included in some thimacs under the assumption the named rectangle indicates its potentiality for creation. Additionally, the events will be represented by their regions. Thus, the static model includes the selected potentialities.

An event is an actual occurrence that involves a region and time+. The plus refers to other factors involved in constructing the event, such as energy. For example, Fig. 6 shows the event *A shark attacks a human being*. For the sake of simplicity, events will be represented by their regions. Fig. 7 shows a sample of dynamic events in this example as follows.

$E_1$: *A certain canary exists.*
$E_2$: *A certain canary sings.*
$E_3$: *A certain canary flies.*
$E_4$: *A certain canary lands after flying.*
$E_5$: *A certain shark exists.*
$E_6$: *A certain human is in the sea.*
$E_7$: *The shark attacks the human.*

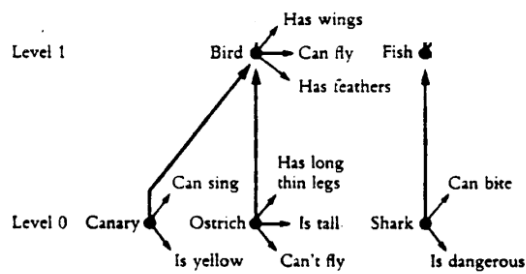

**Fig. 4 Organization of a memory structure of a 3-level (partial from [12]).**

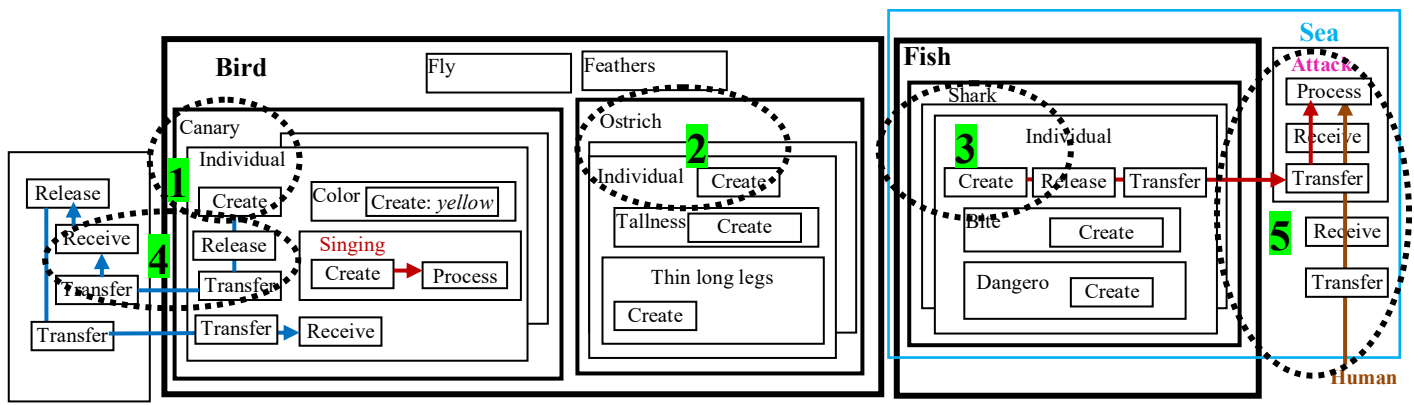

**Fig. 5 TM static model.**

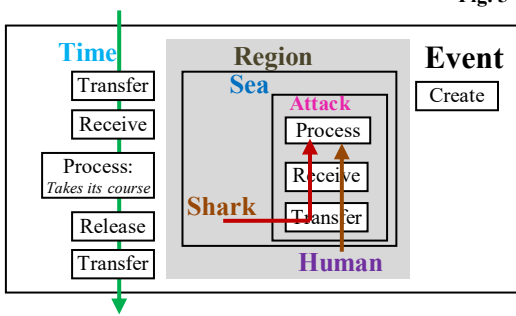

**Fig. 6 The event *A shark attacks a human being*.**

The third canary sings. Simultaneously, a shark appears. A man appears in this world and goes in the sea. An attack occurs involving the shark and the man. Thus, **D**'s history includes all events in their sequence of occurrence.

### B. Example 2: Geographic Information Systems

Modeling and design are key phases in developing systems. A good model helps to achieve modularity and ease of evolution and maintenance. Focusing on GIS, development is intrinsically difficult because of the spatial nature of the information involved in these systems. Location does not change with time, changes in locations might be infrequent, or location might change frequently or continuously, in which case it is necessary to manipulate their evolution.

Suppose that the domain of the world that is represented in this example, denoted as **D**, is initially empty. Fig. 8 shows a sample chronology of events in this domain. First, a canary appears (exists), then a second canary, and then a third canary. The second canary flies in the sky, then lands.

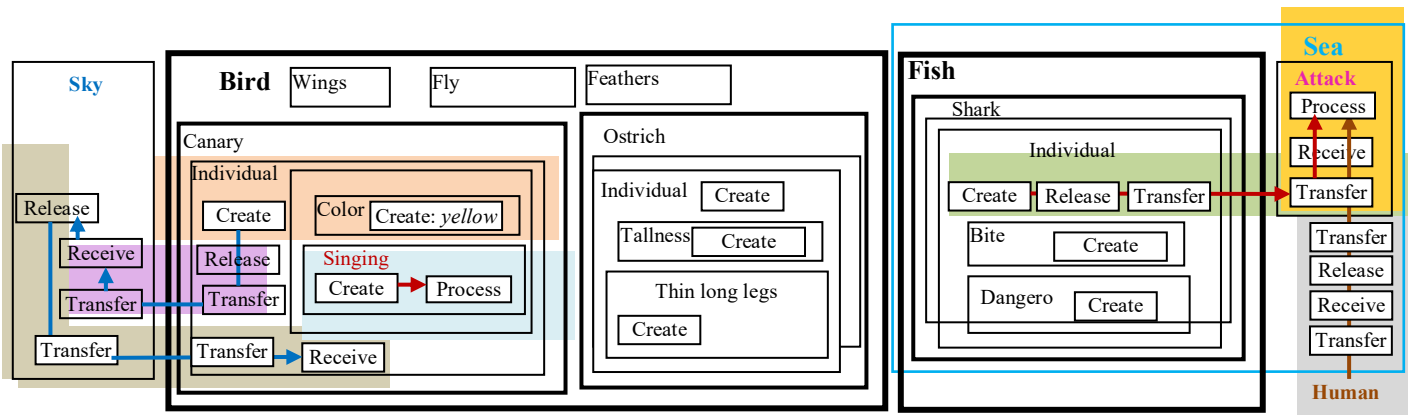

**Fig. 7 Partial TM dynamic model.**

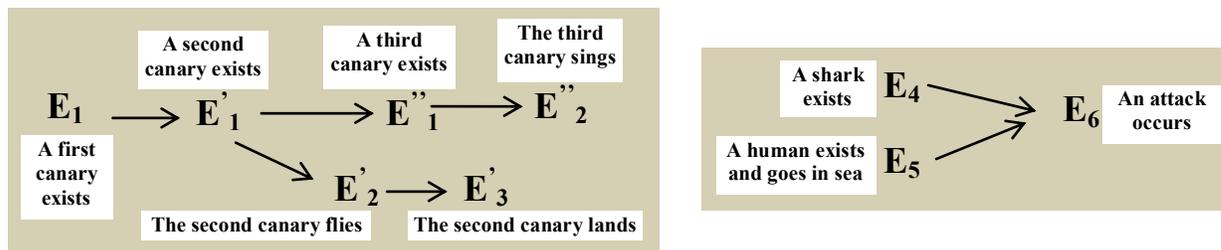

**Fig. 8 Sample chronology of events.**

Representing locations also involves defining operations related to spatial objects: distances, area overlay, the union, and intersection of areas. This task is not so easy; since locations are not merely numbers but are abstract complex objects [13]. Reference [13] discusses a conceptual modeling for GIS using the entity–relationship modeling with its basic components such as entities, relationships, and attributes. An example is given that relates to a city containing land parcels as a principal spatial phenomenon (see Fig. 9). Fig. 10 shows a sample in this context using the popular entity–relationship model.

To conserve space, we model only part of the given land parcels in Fig. 11, which shows the TM modeling (thimacs) of only land parcels: *John*, *Peter*, *Paul*, and *Ann*, in addition to their neighboring streets, *1st* and *2nd*, as indicated in the dotted circle in Fig. 9. The dark boxes in the TM diagram (Fig. 11) represent the land parcels and the portions in red and blue circumferences of rectangles that represent parts of the two streets. All actions (create, process, etc.) are removed for the sake of simplicity, and the two arrows at the top of *John* exemplified further details that can be applied to others. For example, in John, c, e, i, and f are the side lines where only the subthimacs of line c are shown to be described in terms of length and its end points and their coordination (x, y). The street parts are specified in terms of their sides with a fixed width.

The land parcels and two streets are each represented as a thimac, and all types of flows, triggering, and actions can be applied.

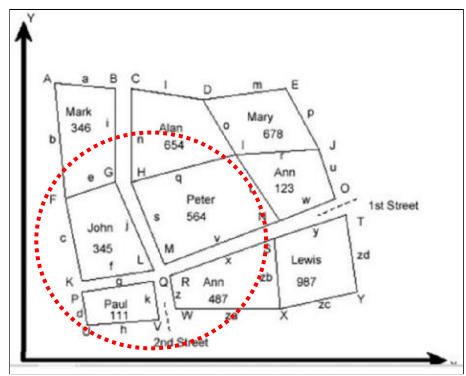

**Fig. 9 An example of land parcels**

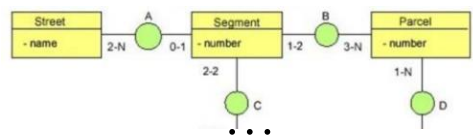

**Fig. 10 The entity–relationship diagram for land parcels (Partial from [13]).**

For example, using owner names as the land parcel name, suppose that (1) PAUL is sold to the new owner DAVID. Then, suppose (2) it is moved to zone 2, and finally (3), its use type is changed to apartments. We focus here on only these three events and add the involved actions and attributes. Fig. 12 shows the model of the three events, $E_1$: *PAUL is sold to DAVID*, $E_2$: *DAVID moves to Zone*, and $E_3$: *DAVID "use type" is changed to apartments.*

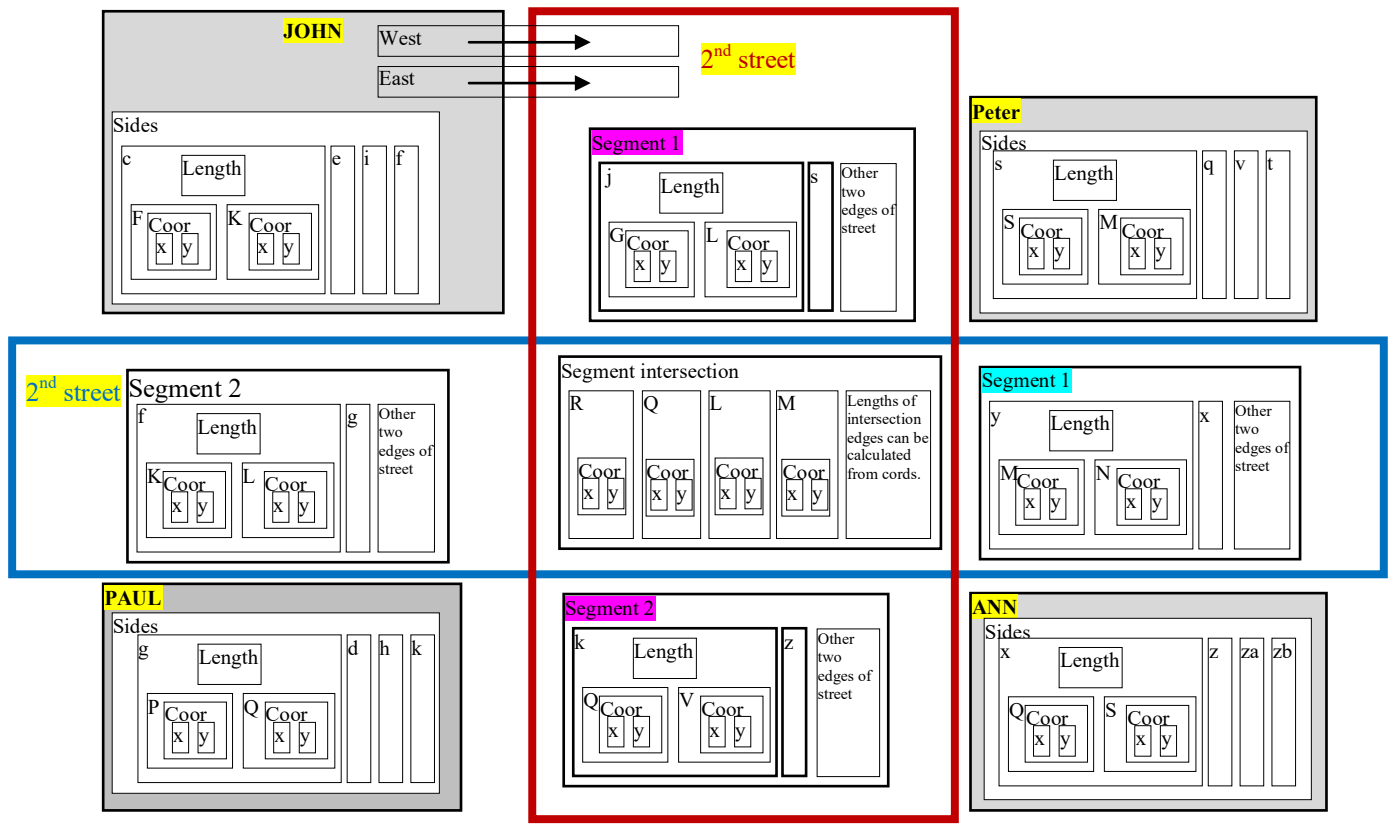

**Fig. 11 The TM model of the land parcels.**

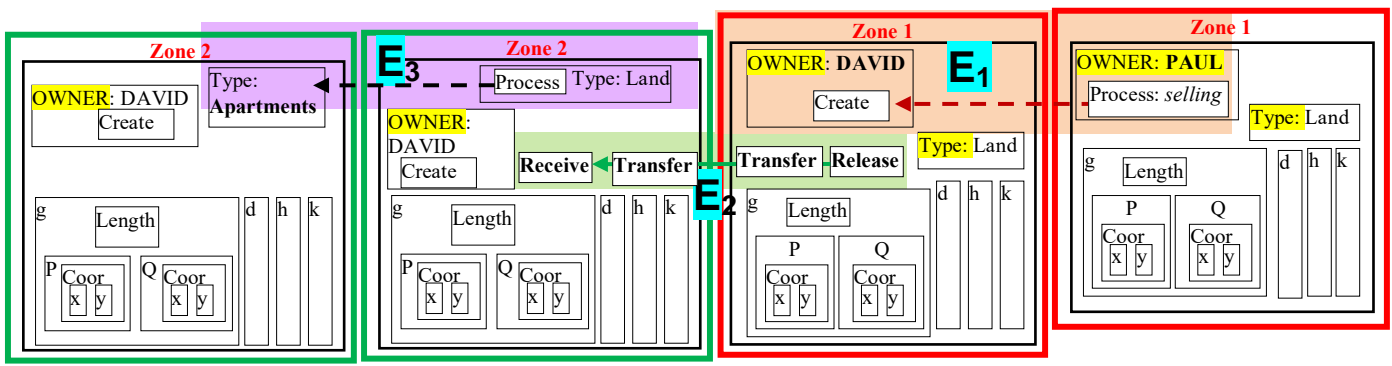

**Fig. 12 Modeling three events: (E₁) *PAUL is sold to David*, (E₂) *DAVID is moved to zone 1*, and (E₃) *The "use type" of DAVID becomes apartments***

### C. Example 3: Zeno and Motion

The Zeno's paradox of motion is given by observing an arrow that has been shot from the bow [19]. One cannot coherently assert that an arrow is actually moving because the arrow needs to be at a certain place at each point in time, which by definition cannot contain any duration at all. However, if this is the case, then the arrow is not moving because all of its trajectories consist of a series of these moments and it is not moving at each moment. So, if it is not moving at one moment, then it is not moving at all [19].

In TM modeling, movement is represented as the flow between two things with a middle interface, as seen in Fig. 13. Note how Aristotle's definition of motion (with respect to place) is divided into TM actions: release, transfer (output), transfer (input), and receive. The intrinsic instability of a "complex whole motion" organizes into four "states" that defuse into instantaneous situations of the moving thing. The four generic actions are now part of a determinate whole of "passing from one place to another."

During movement, generally, there are middles between *from* and *to* things. In fact, there are many *repeated* flows in the middles, as exemplified by three general areas: in Fig. 14, *initial*, *middle*s, and *end*. Each event is a durance in time (i.e., chunks of time, however small) and characterized by continuity. Note that each middle *needs* six generic events (shown in red in Fig. 13; i.e., durations to participate in the movement) so that there is no fear of infinite occurrences of middles.

Suppose that we interpret Zeno's view as "the arrow needs *to be* at a certain *place* at each point of time." Fig. 15 shows the "existing arrow" generic events: an arrow exists in the initial position: released, output, input in the second position, received, etc. Note that the arrow as an object is always being "eventized" (changed in position not in itself); thus, in a dynamic generic region (generic event), we cannot say it is at a certain "*place*" at any point in time of movement. We can say that it is being released, received, etc., at a certain position.

However, such a positioning is not stable because it involves ongoing changes of position. There is no "new existence of the arrow," but there is a flow. In the movement range, for an observer who is watching the arrow, he/she "sees" the continuous change in the position of the *same* existing arrow, not a new position of the arrow.

In the interpretation of Zeno's puzzle, suppose that the arrow is not moving (freezing) because "all of its trajectories consist of a series of these moments and at each moment it is not moving." This is represented in TM, as shown in Fig. 16.

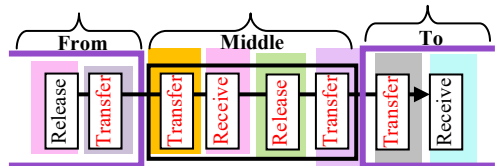

**Fig. 13 Movement from, to and middle.**

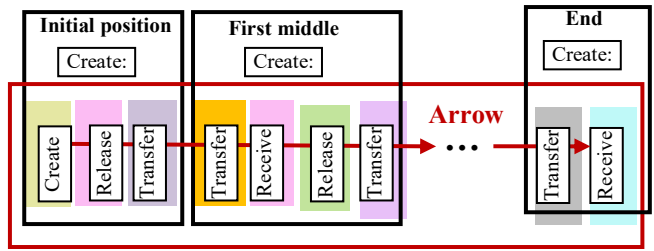

**Fig. 14 Movement of the arrow through positions.**

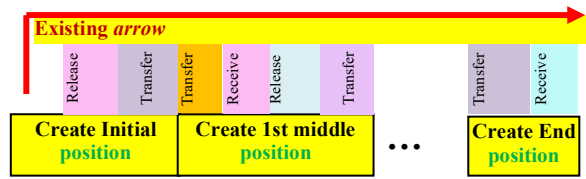

**Fig. 15 Existing arrow flight.**

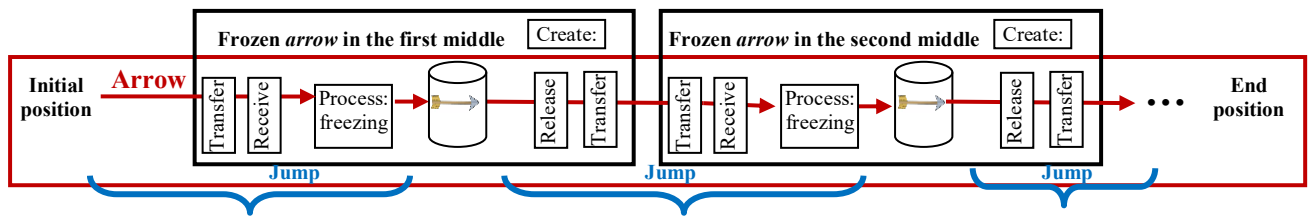

**Fig. 16 The arrow jumps between consecutive positions.**

In this case, the arrow *jumps* from one position to the next one, assuming very fast jumps and minimum stay in each position. Still, these jumps imply movement in contradiction to Zeno's claim of "no movement." Next, we show that the event that we called freezing is a kind of movement.

The arrow cannot be "at a certain *place* at any point of time" because it is continuously changing: being, released, being transfer, being received, etc. The nature of change in generic events of the trajectory of the arrow is illustrated in Fig. 17. The "boundary" between two generic events is an overlapping where the first generic event disappears and the successor event appears. If there is any *freezing*, then it is in the interface between receive and release, as shown in Fig. 17. However, this type of "give and take" is a change or a kind of movement. Thus, the event that we called freezing is movement.

## IV. GENERAL TM-BASED METAPHYSICS

As mentioned in the introduction, the TM model involves two modes of being, dynamic *existence* and static *subsistence*, that represent a targeted portion of reality. *Something-ness* is either subsisting or existing. Thingness is a reflection of thimacness; everything is a thimac. Nothing-ness is *non-thimacness* (i.e., outside reality: nonsubsistence) and nonexistence (i.e., no region and no event).

The next two sections, 5 and 6, center on the subsistence of regions and the distinctiveness of events, respectively. Specifically, in these sections, we make two claims: (i) the perceptibility of regions as a phenomenon; and (ii) the distinctiveness of existence (pure existence) as a vehicle for events.

In preparation for such assertions, this section includes discussions about the nature of thingness and thimacness and the fact that subsistence does not have a transcendental nature because subsistence and existence are *superimposed* over each other in reality.

### A. Thingness as Thimacness

TM modeling describes reality in two levels of a potentiality/actuality scheme adopting an idea that goes back to the Stoic modes of reality (see Fig. 18). A two-level reality is an old idea. According to traditional interpretations, there is the doctrine of degrees of reality in Plato's philosophy. The doctrine of degrees of reality says that forms (knowledge: e.g., *oddness of three*) exist while particulars (opinions, e.g., *Simmias is not very dependable*) are half existent and half nonexistent [20]. Russell held that universals (e.g., chairs or tigers) do not exist; they *subsist* and are nonetheless "something" [21]. According to Russell, subsistence is opposed to "existence" as being timeless.

The two-level depiction is made to emphasize and illustrate the characteristics of each of the two levels. In TM modeling, the subsistence does not have a transcendental nature because subsistence and existence are *superimposed* over each other in reality. Subsistence is a mode of reality where we consider a thing apart from its existence (e.g., squareness). This subsistence is the double of existence. For example, Russell asserted that it is unnecessary to assume the *existence* of negative facts. In TM, negativity has a subordinate ontological status. The negative of an occurrence is "no event" [22].

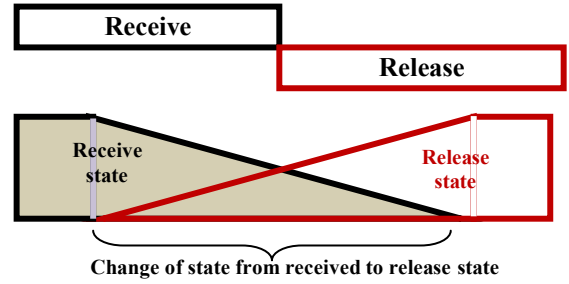

**Fig. 17 The boundary between two successive events.**

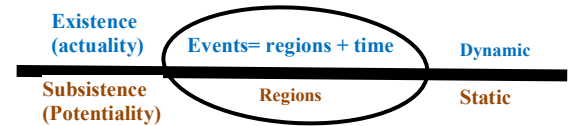

**Fig. 18 Two-level TM modeling.**

The negation of *Something exists* is *Something subsists*. *Something subsists* is either a static **region** that may later exist (the so-called synthetic proposition), or it is not mappable to existence (e.g., square circle). In this case, we use Lupasco's "no event" (see [22] to claim that "no event" reverts from the dynamic level to the static level (i.e., the event returns to potentiality to become only a region).

### B. Subsistence in Existence

Therefore, existence and subsistence are like a double-image impression (e.g., Rubin's vase), which is possible with a figure-ground perception. They present in reality an analogy of *universal* that appears "in the many" particulars (events). Every existence (event) contains subsistence (region) and both (event and its region) are in *existence containers* (to be discussed in the next section). This is in line with Russell's account that the components of the world are actual and possible sense data [23]: *subsisting things can be sensed in existing things.* The psychological subsistence is the region of previous existence.

The TM static level is a world without time. This implies the *simultaneous presence* of all things in the universe. As we will discuss later, this is not necessarily true because there is no separate "universe": one for regions and one for events. The subsistence of regions resides in existence (of events). Time is immaterial to the regions definition because they are *hiding* within events. There is no simultaneous appearance of regions *by themselves* because regions can be those in the history of the universe. This does not contradict the creation of new regions from previous ones.

## V. REGIONS OF THE SUBSISTENCE LEVEL

As discussed in a previous paper [24], subsisting things in reality originate from *first occurrences* in reality. Thus, regions are not found independently of events, just as processes cannot exist without their events (e.g., traffic embedded into existing cars, roads, and lights). The big claim in this study is that regions are *perceivable phenomena*.

### A. First Occurrences of Things in Reality

The question of how things emerge where there were no things before is a very old philosophical question. In TM, subsisting things in reality (regions) originate from their first occurrences (events) in reality. The first event was not included in reality prior to the act of becoming. According to [25], there was a reality when even the sun was new. For example, the thimac *Grass as food* is a new thimac in reality after the appearance of the first organism (thimac) that feeds on grass. (This example is from [25].)

Physics recognizes many first-time creations. For example, at about .0001 seconds after the Big Bang, "Quarks, for the first time, can combine in groups of two and three to become neutrons, protons and other types of heavy particles" [26]. At the end of the "first three minutes," we find an event that provides us with the nucleosynthesis of the light chemical elements, such as deuterium, helium, and lithium [26].

A thimac born in existence establishes a foothold in the world through subsistence. Additionally, this process involves the formation of compound, higher-order thimacs out of simpler, lower ones. This process is similar to induction that "proceeds through an enumeration of all the cases" (Aristotle) to arrive at the higher categories by means of abstraction. In Whitehead's words, in *Process and Reality*, "the many become one and are increased by one."

In TM, the abstraction is oriented toward the gestalt of the entire scene (TM subdiagram) of events as a whole thimac that is supplied with generic generalized actions. The process also involves induction to derive a whole from the parts.

The new generated thimac inherits the same thing/machine structure. The new event establishes a new region. In its reproduction as an event, it evolves with variation (e.g., subthimacs) that can be generated in agreement with the so-called process of Darwinism, where it is assumed that thimacs develop through variation and selection and transmission.

The detachment of regions from events is the opposite of the emergence of events from regions into existence. After the first emergence, each new existence (event) involves the lower stage of reality: subsistence. For example, life emerged in terms of the physicochemical process that forms a complex material body and mind of a living thing. The *emergent* quality is the summing together into a new totality of the component materials. (This description of life emergence is taken from [27].) What we call detachment refers to the operation of extracting a region from an event. The extraction process is in Husserlian language grasping the region of an event in "one glance." Different types of thimacs (only create, create+process, release+transfer, etc.) are "units" of wholes. They fit together through flows to form a unified whole just like adjacent pieces in a jigsaw.

The argument here is that humans have ability (maybe with variable degree) of detaching its region from an event. There is no suggestion here that such a region is in the *form* of TM diagrammatic language.

### B. Region Unity

A region is the essence of its event [24]. The following discussion aims at applying the notion of *unity* to both a region and event. We imply that a region is a whole, just as an event is a whole in preparation to claim the perceivably of regions. Note that the purpose of binding regions and events in existence is to reject that regions have transcendental nature, thus avoiding the objectionable type of metaphysics in conceptual modeling.

An *entity* (TM event) in the classical sense is a thimac in which the interior subthimacs are held together by "dynamical bonds" (flows) between them, which have the effect of individuating the thimac as a whole from its environment. Internal subthimacs work together to ensure a "cohesive system (regions or events)." At the TM dynamic level, the internal bonds constrain the behavior of its constituent sub-thimacs in such a way that the totality behaves dynamically as an integral whole. (This description is adopted from [28] and applied to TM.) According to [28], "Wherever we find such a system, we are able to *identify* and *re-identify* it. That is, what licenses our calling it an entity. Entities are not basic; they are derived and entities are certain kinds of persistent, cohesive processes" (italics added). Cohesive systems (i.e., entities) are typically manifest properties that are different from those of their internal subprocesses. Some holistic properties (subthimacs) may result from an aggregation of the properties of the constituent thimacs, thus providing the sense in which the main thimac can be said to be "more than the sum of its subthimacs."

### C. Region Perceivability

When we see an event, we simultaneously *perceive its region*, as illustrated in Fig. 19. The region perception is a reduction process of the complex of thimacs in an event to a minimum coherence. We seem to apprehend both at once. Such a phenomenon is facilitated by having the region (essence) as a part of the event. In TM, regions are entirely interior to events, but nothing that happens to an event can alter its region.

Deleuze gives an example of a similar type of perception that extracts structure from events. According to Deleuze, "I see a cube, but I cannot see more than three of its sides at any given moment. However, I can extract the cube from where I find it and proceed to manipulate it."

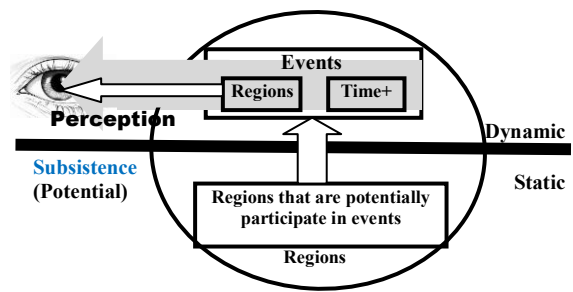

**Fig. 19 Regions subsists in events.**

From this, Deleuze concludes the following about entities: "the cube must have a private, internal ground in and of itself... There must be something by virtue of which it remains *this* cube throughout different settings" [29]. In TM, a perceived existent (existing thing) is implanted *with* a TM region.

Thus, subsisting things are not far removed from sense experience. Subsistence is in existence just like a white square floating weightlessly in a white field: an abstraction without reference to external reality [30]. It seems that this requires a labor of abstraction to achieve a type of higher-order perceptual sense of integrated (static) structure.

In TM, regions of events can be found from being components of (abstracted from) their events. However, the static region does not exist in (temporal) actuality by itself. It is but one ingredient of the event. Note that everything in the TM world has an individual region independent of the thing that has it, whether the thing is actual (object or process) or static (non-actualized). Additionally, dependencies among things (e.g., existentially dependent) are specified in the static level (e.g., region A *triggers* the creation of region B).

The region is determined by its "net" of the interconnected TM actions that work together to form the conceptual scene of event. This process is, roughly, the opposite of the so-called "graph perception" where people decode information represented in a graph. The process involves perceiving graph forms (thimac regions) from events to capture wholes. According to Hume, "whatever we conceive, we conceive to be existent." This implies for us that, at least, the region is an important side of this perception. Regions are sensible (i.e., capable of being perceived). They are limited to an inductive level of knowledge. In this context, perception signifies gaining awareness of hidden regions (forms) within events through the senses.

It seems that humans abstract events into TM regions (awareness chunks not necessarily in a diagram form) through processed sensory data of senses: see, touch, hear, smell, and taste. Some philosophers view the analysis of such perception as an analysis of the *common sense* to arrive at three types of "sense impressions":

(1) There is (create) a thing,
(2) Variation/change of a thing (process), and
(3) Movement of a thing (release, transfer, and receive) from where to where.

Such a process is illustrated in Fig. 20. The event itself embeds its timing; thus, staticity is assumed as a default.

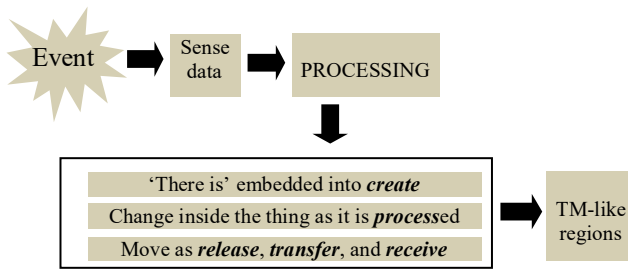

**Fig. 20 An event is abstracted into three types of manifestations.**

For example, when people are asked (e.g., by a newsman) about "what happened" in an event (this afternoon), they focus on three aspects: *there is* something, altering aspect (laughing, threatening), and moving (e.g., running, driving). In this approach, the holistic grasp of a world event is closer to being a series of pictures.

### D. Abstraction

The type of abstraction of the perceived scenes involved is supposed to somehow strip them in order to free them from their "existence" features and thereby make them a static region. Perception here refers to various types of information obtained about some event of the external world. It is abstraction of form that is flattened from any content. The form informs a certain unity to complex things (e.g., TM subdiagrams) that are directly perceived by spectators. The form in the TM contexts includes entities (things), actions (machines), and forms (regions).

In this view, we perceive immediate subsisting reality (based on generic actions; i.e., create, process, release, transfer, and receive) when grasping existing reality. The subsisting perception of a region's wholeness is awareness of the "being" (of a low-resolution something), which forms the nucleus of perceiving existence (of a high-resolution dynamic thing). This awareness is both experienced and learned as we use past experience, organized as subsisting *regions*, to guide our construction of existing events.

Aristotle spoke of "some one thing" a faculty called *koine aisthesis* (the mental faculty of common perception, e.g., shape, which is peculiar to no one sensory modality [31]) and its role in the perception of objects "to coordinate, simultaneously, the reception of qualities in a thing that impinge upon the individual senses" [32] and to reflect a "common sensibility to particular features" [31]. As Aristotle observes, the activity of differentiation or discrimination by which any two qualities (in our case, perceiving TM staticity and dynamism) can be brought into relation for sensibility necessitates a simultaneous awareness of this difference within sensibility [32]. In modern philosophical language, we elicit from the *particulars* (TM events) *generalities* (TM regions, e.g., essences.). Aristotle viewed such a faculty as "a unity divided only into what he calls 'conceptual parts' ... or 'logical parts'" [31]. Being a conceptual part of a thing is merely being one of its TM subregions. From the TM point of view in this context, the capacities and elements of a thing (static region) can *subsist* without the thing existing. Not every static region (thing) necessarily obtains existence. Thus, the realm of events is narrower than the static realm.

Aristotle presented an analogy to make this formulation more comprehensible: that of the *point* and the *line* [32]. One definition of a line is as follows: a line is a set of collinear (lying in the same straight line) points that are connected in a one-dimensional plane. Fig. 21 shows the static region that corresponds with an existing line. The static region is embedded into existence, as explained in the figure.

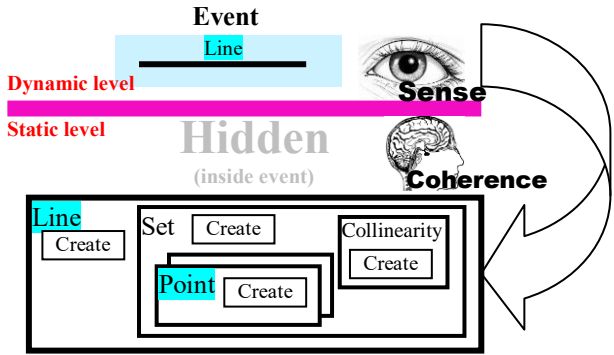

Fig. 21 A static region can be sensed from into its event.

## VI. EXISTENCE IN THE EVENTS LEVEL

With this understanding of the TM region, in this section, we focus on existence (pure existence) as a vehicle for events.

### A. Existence

Generally, the notion of existence encompasses questions related to the concept of *existence at large* where the TM model defines events as "what it means to be an existing thing." An event is *what happens* in a region. According to [21], we can understand the Platonistic intuition wherein universals and individual things have different modes of being *without implying that there is more than one fundamental concept of being*. Analogously, TM existence and subsistence have different modes of being without implying that there is more than one fundamental concept of reality.

Reality is viewed as a composite thimac with a singular unified totality. The notion of region is similar to Lynne Baker's [33] *constitution*. According to [33], suppose that a lute, one of the musical instruments, was smashed and whose only remains were slivers. Did anything really go out of existence when the lute was smashed? According to [33], there are three possibilities:

(1) No lute ever existed. All that existed were slivers arranged lute-wise. When the lute was destroyed, the only change was in the arrangement of the particles.

(2) Lutes are really just the matter (extension) that occupies certain spatial or spacetime points. Lutes are identical to sums of particles. Any matter-filled spatial or spacetime points have sums. We give names for some of the sums that are arranged in certain ways (e.g., "lutes").

(3) Lutes really exist in their own right; they are irreducible to anything more basic. Particles made up the lute that was smashed, but the lute was not just identical to particles arranged lute-wise.

According to [33], if lutes are not just identical to sums of particles, constitution is the relation between the lutes and the aggregates of particles that made them up. Constitution is "a relation that holds between granite slabs and war memorials, between sodium and chlorine atoms and salt molecules, between pieces of paper and dollar bills—things of basically different kinds that are spatially coincident."

### B. The Exicon

In TM modeling, such a constitution is represented as a subsisting region that is actualized as an event. Fig. 22 shows the TM model that corresponds to the lute-smashing scenario. To save space, only the dynamic TM model is shown. First, the lute, including its slivers, exists ($E_1$) to be smashed ($E_2$) in order to produce a pile of slivers.

Let us focus on the *create* action of the lute as shown in Fig. 23. Fig. 23 expresses Event 1: *A created lute* or *A particular lute exists*. Fig. 24 shows another representation of the event in terms of a region and time. Fig. 25 shows that the event by itself is a slice of existence of *any* region in time. A slice of existence can be abstracted from any region. This slice of existence will be called an *existence container* (exicon). Fig. 26 shows an exicon as a part of existence. Event(s) occupy exicons (individual existences) that "stand under" or uphold events throughout existence as a *phenomenal* trait.

The exicon is a TM "non-thing" agent that is responsible for the existence mode of events. It is a slice of existence by itself or what it is to be existent analogous to the Aristotlian *soul* as what it is to be alive. There are numerous exicons coming into existence and going out of the realm of existence everywhere all the time. An exicon can contain any event.

A deeper observation for the notion of existence is that regions, the very essence of thingness, are accidents of exicons. That is, regions are accidental to exicons as pure existents. Note the exicon is not accessible to human consciousness and is grasped only in the setting of an event.

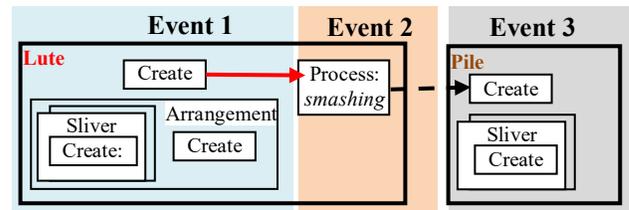

Fig. 22 Smashing the lute produces a pile of slivers.

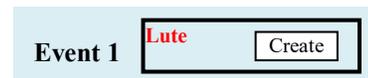

Fig. 23 A particular lute exists.

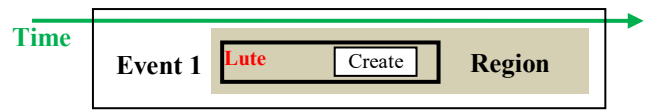

Fig. 24 Another representation of an event.

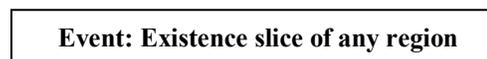

Fig. 25 Existence slice of any region.

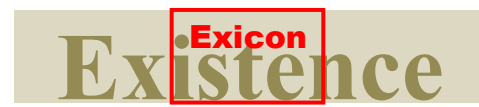

Fig. 26 Capsule of existence.

Exicons give the TM modeling a sense of a kind of depth, thus providing a reasonable level of feeling or showing hopefulness in TM representation in comparison to imported, chewed-over ontological notions (e.g., class, object) from philosophy.

### C.  Deleuze and Exicon

To gain some level of historicity to the notion of exicon, we may try to link it to Deleuze's "everything is a machine" mentioned previously. According to Deleuze, "whenever machines [*TM machines*] combine their forces [*TM trigger*] to produce a water molecule, a marriage, a perception, a house, or a red panda, **a body without organs** [exicon] emerges" [29]. Note that, contrary to Deleuze, in TM, the *body without organs* is not a machine; it is an exicon. In TM, a *body without organs* does not have the *create*, *process*, *release*, *transfer*, or *receive* actions.

In Deleuze's famous example, an actual egg is a TM event that emerges as the fertilization of an egg region with an exicon. Deleuze also asserts that a body without organs is autarchic (absolutely self-contained). A machine, in TM, is what "produce(s) or fabricate(s)" (these are two Deleuzian terms) or *events' generator*, and "it does not preexist…Whenever and wherever an entity [region] comes into existence, it immediately has its body without organs [exicon] as the guarantee of its irreducibility" (Deleuze) [29].

Hereafter, we will use Deleuze's body without organs and exicons, interchangeably ignoring that an exicon is not a machine. In the following discussion, we select some of the Deleuze's ideas and apply them in the TM exicon. The exicon never manifests directly in events. It is always enclosed and covered up by its own actual manifestations, from which it differs in kind. The perception of exicons is like black holes that cannot be perceived directly so that their persistence must be inferred from their events. A football, for example, has subthimacs (leather, air on its inside and outside, and flows to/from hands and feet), but its exicon never coincides with these things.

### D.  Further Exploration of Exicons

Thus, the claim in TM is that existence is unthinkable without the notion of an exicon. The exicom makes a region+ capable of existing. The "+" refers to other elements involving materiality, and so forth; it may be called entelechy of things, as the process of actualizing potential things.

The exicon stands in its own right as a *non-thimac* (no create, no process, no release, no transfer, and no receive), even though, empirically speaking, such an exicon still "participates" in existence of things. Accordingly, an event sliced in two in (1) a region, reveals itself, and (2), in an exicon, it conceals itself. It is unknown how it is *to bring together existential dependence between a region and an exicon*. This is not an uncommon idea. According to [34], "in modern physics it is not clear how concepts such as field, energy, and force could receive adequate materialist interpretations."

The exicon possesses a quality that causes the igniting and persistence of existence instantiation of the region along the exicon's "life." However, one thing seems to "carry out" in this context entropy or decay of events. An event "ceases to exist" means the disintegration of its "existence wholeness"; hence, its exicon drops away. The event itself is a thing (region) and "no thing" (exicon). Physics has shown that "nothing" actually represented where something could exist. "Nothing" is speculated to be a stage of the universe's existence. In the *Casimir effect* case, "Ask a physicist about a vacuum, the very definition of nothingness for most of us, and they will tell you it is pulsing with activity. According to quantum theory, in a vacuum wave-like fields are constantly fluctuating, producing particles and their antimatter equivalents that fizzle in and out of existence" [35]. The *non-thimac* is not a kind of being (i.e., not a thing), and it is not a machine. The ancient Greek *Arche* may be the closest term to exicon.

Existence is, essentially, a machine that is enveloped into exicons.

- An exicon is a pure plane upon which unformed events wait for their regions.
- Exicons come in all sizes to "fit" all kinds of machines. Both Pinocchio's wooden region and Pinocchio's human region have exicons.
- Exicons form a "space" of preexistence where they do not exist and move to an actuality (events) arena without being fertilized with regions.
- Axicons are there without exhaustion and do not ever terminate, functioning as exterior shells being emptied and loaded with regions.
- Axicons represent a unity in existence because the "individual" exicon dissolves back into pure existence after departing its event. On the other hand, events represent multiplicity in ordinary (phenomenal) existence.

### E.  Preexistent and Nothingness

An exicon is an instance of preexistence and time of "indeterminate states" (unactualized time) before the formation of an event. Accordingly, exicons can be called *preexistents*. Fig. 27 shows the relationship relationships in the context of this discussion. *Preexistent* is a very old notion. Plato described preexistence as emanation: the rays of the sun are to the earth. Hegel described preexistence as the place where "no thing can be distinguished from another." For Nietzsche, it is the eliminations of all distinctions.

From a Heideggerian prospective, existence implies "the manifestation of a Being through Entity" [36]. Being, here, has some of the aspects of an exicon. According to Heidegger, the high school building stands as an entity; on the other hand "Being does not consist in our observing entity…One can, as it were, smell the Being of such buildings, and often after decades one still has the scent in one's nose…the *Being* is only matched by the *non-Being*" [36].

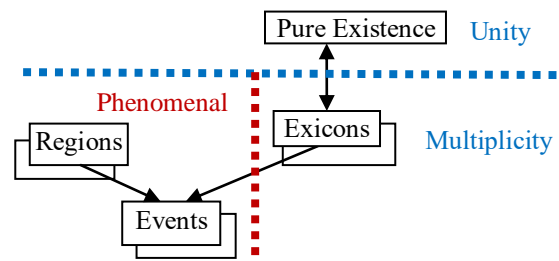

Fig. 27 Relationships between pure existence and phenomenal existence

Similarly, in "the quantum mechanical language, upon measurement, a wavefunction…in the language of metaphysics, bring[s] a being into existence collapses it into entity…the metaphysical being cannot be physically reached but only grasped through its entity manifestation" [36].

Using Deleuze's language *A machine does not "whirr" (Deleuzeian word), i.e., event-ized, without exicons*. In Deleuze's example, the breast, milk, and the mouth form only a *region* of a composite machine at TM, the static level. No milk would reach the mouth, just like no water runs in a map. Events emerge when permeating the regions with exicons.

Destroying an exicon can be illustrated in the TM model of the lute example; first, the lute exists, and it is smashed to become a pile of slivers while the lute region returns to the subsistence level to be actualized as a pile. The events are potentialities that become actualities according to the chronology of events. Saying that lutes went out of existence refers to the subsistence of the lute. The smashing of the lute was the loss of Event 1 (the lute itself). According to [33], "the smashing of the lute was a genuine loss to reality—an ontological loss. In TM, it is an "occupied holistic" exicon destroyed to some of its sub-exicons. It is a loss in the sense of losing "wholeness" and a gain to entropy/destruction.

## F. Exicon and Substance

Exicon may be related to the notion of substance in philosophy. Often, substance is described as what "stands under or grounds [existing] things" [37]. The notion of exicon clarifies the notion of substance. For example, "for an atomist, atoms are the substances, for they are the basic things from which everything is constructed" [37]. In TM, there are three strains of an atom: region, event, and exicon. The region is the blueprint, form, or essence of an atom (e.g., nucleus of protons and neutrons with electrons in orbits). An atom as an event is an actual tangible entity. The exicon is a slice of existence that holds the atom steadily in existence.

For Kant, substances are those enduring particulars that give unity to our spatio-temporal framework, as well as the individuation and redentification, which enables us to locate ourselves in that framework [37]. In TM, exicons are not particulars but constituents of particulars in addition to regions+.

The standard four-dimension metaphysics [33] views concrete objects as spacetime worms that have temporal parts as well as spatial parts. According to [33], "There are countless, nameless spacetime worms coming into existence and going out of existence everywhere all the time. With the super-abundance of worms beginning and ending at every spacetime point." In TM, the temporal parts of these warms are exicons.

## VII. Conclusion

This paper has introduced an exploration of the metaphysical origins of the preconceptual modeling phase of software engineering (Fig. 1), as exemplified by the high-level model called TMs. We employed the so-called TM modeling that is based on the notion called *thimac* (things/machine). The TM modeling has pursued a broader metaphysical study of the nature of the domain's things and processes beyond categorization. Discussions involve the nature of thimacs (things and processes) and subsistence and existence as they are *superimposed* over each other in reality. The results contribute to further overarching understanding of TM modeling, in addition to introducing some metaphysical insights. Future research will pursue further exploration about the nature of thimacs.

## References

[1] S. Vössner, "Pre-conceptual modeling for exploring actors and interactions in real world sytems," *NEMO Summerschool Lecture*, July 26, 2023, https://nemo.omilab.org/nemo/2023/lectures/36-voessner.html.

[2] P. Haumer, M. Jarke, K. Pohl, and K. Weidenhaupt, "Improving reviews of conceptual models by extended traceability to captured system usage," *Interact Comput.*, vol. 13, no. 1, September 2000, pp. 77-95.

[3] R. N. Memon, S. S. Salim, and R. Ahmad, "Identifying research gaps in requirements engineering education: an analysis of a conceptual model and survey results," *2012 IEEE Conference on Open Systems*, Lumpur, Malaysia, October 21-24, 2012.

[4] Y. Wand and R. Weber, "Mario Bunge's ontology as a formal foundation for information systems concepts," in Studies on Mario Bunge's Treatise, P. Weingartner and G. J. W. Dorn, Eds. Atlanta: Rodopi, 1990.

[5] R. Lukyanenko, B. M. Samuel, J. Parsons, V. C. Storey, and O. Pastor, "Datish: a universal conceptual modeling language to model anything by anyone," *ER2023: Companion Proceedings of the 42nd International Conference on Conceptual Modeling: ER Forum, 7th SCME*, Lisbon, Portugal, November 06-09, 2023.

[6] K. Engelhard, "Inductive metaphysics, editors' introduction," *Grazer Philosophische Studie*, vol. 98, 2021, pp. 1-26.

[7] C. M. Zapata, O. Pastor, G. Guizzardi, and R. Guizzardi, "Concept categorization in pre-conceptual schemas," *Conferencia Iberoamericana de Software Engineering*, 2011, https://api.semanticscholar.org/CorpusID:52833053.

[8] J. Insuasti, F. Roa, and C. M. Zapata-Jaramillo, "Computers' interpretations of knowledge representation using pre-conceptual schemas: an approach based on the BERT and Llama 2-Chat models big data and cognitive computing," Big *Data and Cognitive Computing* vol. 7, no. 4, 10.3390/bdcc7040182.

[9] S. Al-Fedaghi, "UML sequence diagram: an alternative model," *Int. J. Adv. Comput. Sci.*, vol. XII, no. 5, 2021, pp. 635-645.

[10] S. Al-Fedaghi, "UML modeling to TM modeling and back," *Int. J. Netw. Secur.*, vol. XXI, no. 1, 2021, pp. 1-13.

[11] S. Al-Fedaghi and O. Alsumait, "Towards a conceptual foundation for physical security: case study of an IT department," *Int. J. Saf. Secur. Eng.*, vol. 9, no. 2, 2019, pp. 137-156.

[12] A. M. Collins and M. R. Quillian, "Retrieval time from semantic memory," *J. Verbal Learning Verbal Behavr.*, vol. 8, no. 2, April 1969, pp. 240-247.

[13] S. Gordillo and R. Laurini, "Conceptual modeling of geographic applications," in Advanced Geographic Information Systems – Vol. I - Conceptual Modeling of Geographic Applications, S. Gordillo and R. Laurini, Eds. EOLSS Publications, September 2009.

[14] C. Murphy, "Existence-based access the future of cybersecurity," January 5, 2019, https://www.linkedin.com/pulse/existence-based-cybersecurity-christopher-murphy.

[15] C. De Koninck, "Metaphysics and the interpretation of words," *Laval Théologique et Philosophique*, vol. 17, no. 1, 1961, pp. 22-34, https://doi.org/10.7202/1020000ar.

[16] A. S. Kleinherenbrink, *Machine Philosophy Gilles Deleuze and the Externality of Entities*, Nijmegen: RU Radboud Universiteit, November 30, 2016. https://repository.ubn.ru.nl/handle/2066/161709.

[17] M. Heidegger, "The thing," in Poetry, Language, Thought, A. Hofstadter, Trans. New York: Harper & Row, 1975, pp. 161-184.

[18] D. Liebesman and R. K. Sterken, "Generics and the metaphysics of kinds," *Philos. Compass*, vol. 16, no. 7, e12754, https://doi.org/10.1111/phc3.12754.

[19] S. Hongladarom, "Metaphysics of change and continuity: exactly what is changing and what gets continued?" *J. Philos.*, vol. 2, May 2015, pp. 41-60, doi: 10.5840/kilikya2015229.

[20] G. Vlastos, "Degrees of reality in Plato," in New Essays on Plato and Aristotle, R. Bambrough, Ed., Humanities Press, 1965, pp. 1-19.

[21] A. Cusmariu, "Subsistence demystified," *Auslegung: A Journal of Philosophy*, vol. 6, no. 1, 1978, pp. 24-28, https://doi.org/10.17161/AJP.1808.8926.

[22] S. Al-Fedaghi, "Lupascian non-negativity applied to conceptual modeling: alternating static potentiality and dynamic actuality," October 27, 2022, arXiv:2210.15406.

[23] G. Couvalis, *Aristotle on Being: An Aristotelian Critique of Russell's Theory of Existence, 2015 - Modern Greek Studies*, Australia and New Zealand: 2015, pp. 41-50.

[24] S. Al-Fedaghi, *Exploring Conceptual Modeling Metaphysics: Existence Containers, Leibniz's Monads and Avicenna's Essence*, Location: Publisher, February 2024, hal-04469798

[25] C. Hartshorne, *Creativity in American Philosophy*, Albany: State University of New York Press, 1984.

[26] Astronomy, Physics, "Before the Big Bang," July 29, 2017, https://sten.astronomycafe.net/tag/planck-era/.

[27] S. Alexander, *Space, Time and Deity*, London: Palgrave Macmillan, 1966, pp. xx-xx.

[28] R. Campbell, "A process-based model for an interactive ontology," *Synthese*, vol. 166, 2009, pp. 453-477.

[29] A. Kleinherenbrink, *Against Continuity*, Edinburgh: Edinburgh University Press Ltd., 2019.

[30] K. Malevich, *Suprematist Composition: White on White*, MoMA, 1918. https://www.moma.org/collection/works/80385.

[31] P. Gregoric, *Aristotle on the Common Sense*, Oxford: Oxford University Press, 2007. https://ndpr.nd.edu/reviews/aristotle-on-the-common-sense/.

[32] R. Drake, "Aristotelian aisthesis and the violence of suprematism," *Epoché*, vol. 18, no. 1, Fall 2013.

[33] L. R. Baker, "A metaphysics of ordinary things and why we need it," *Philos.*, vol. 83, no. 1, 2008, pp. 5-24.

[34] C. T. Wolfe, *Materialism: A Historico-Philosophical Introduction*, Dordrecht: Springer, 2016, pp. xx-xx.

[35] J. Howgego, "Can we get energy from nothing?" *New Sci.*, September 2, 2015. https://www.newscientist.com/article/mg22730370-800-can-we-get-energy-from-nothing/.

[36] A. M. V. Brânzanic and R. Silaghi-Dumitrescu, "Quale mechanics as a metaphysical weltanschauung of quantum mechanics," *History of Science*, vol. 13, no. 1, June 2023, pp. 10-33.

[37] H. Robinson, "Substance," in The Stanford Encyclopedia of Philosophy, E. N. Zalta, Ed., Fall 2021. https://plato.stanford.edu/archives/fall2021/entries/substance/.